\documentclass[manuscript=article,journal=jpclcd]{achemso}
\usepackage{amsmath}
\usepackage{graphicx}
\usepackage{amssymb}
\usepackage[usenames]{color}
\usepackage{subfigure}
\usepackage{dcolumn}
\usepackage{bm}
\usepackage{ulem}
\usepackage{color}
\renewcommand{\eqref}[1]{(\ref{#1})}
\bibstyle{achemso}
\setkeys{acs}{usetitle = true}

\makeatletter
\title{Direct Measurement of Competing Quantum Effects on the Kinetic Energy of Heavy Water upon Melting}
\author{Giovanni Romanelli}
\affiliation {Universit\`{a} degli Studi di Roma "Tor Vergata", Dipartimento di Fisica e Centro NAST, Via della Ricerca Scientifica 1, 00133 Roma, I}
\author{Michele Ceriotti}
\email{michele.ceriotti@chem.ox.ac.uk}
\affiliation {Physical and Theoretical Chemistry Laboratory, 
University of Oxford, South Parks Road, Oxford OX1 3QZ, UK}
\author{David E. Manolopoulos}
\affiliation {Physical and Theoretical Chemistry Laboratory, 
University of Oxford, South Parks Road, Oxford OX1 3QZ, UK}
\author{Claudia Pantalei}
\affiliation {Laboratoire L{\'e}on Brillouin, CEA, Saclay, F}
\author{Roberto Senesi}
\email{roberto.senesi@uniroma2.it}
\affiliation {Universit\`{a} degli Studi di Roma "Tor Vergata", Dipartimento di Fisica e Centro NAST, Via della Ricerca Scientifica 1, 00133 Roma, I}
\affiliation {Consiglio Nazionale delle Ricerche, CNR-IPCF, Sezione di Messina, I}
\author{Carla Andreani}
\affiliation {Universit\`{a} degli Studi di Roma "Tor Vergata", Dipartimento di Fisica e Centro NAST, Via della Ricerca Scientifica 1, 00133 Roma, I}

\begin{document}
\singlespacing

\newcommand{\avg}[1]{\ensuremath{\left<#1\right>}}
\newcommand{\remove}[1]{}
\newcommand{\oxford}[1]{#1}
\newcommand{\rome}[1]{#1}

\begin{abstract} 
Even at room temperature, quantum mechanics plays a major role in determining the 
quantitative behaviour of light nuclei, changing significantly the values of 
physical properties such as the heat capacity.
However, other observables appear to be only weakly affected by nuclear quantum 
effects (NQEs): for instance, the melting temperatures of light and heavy water 
differ by less than 4~K. 
Recent theoretical work has attributed this to a competition between intra and inter
molecular NQEs, which can be separated by computing
the anisotropy of the quantum kinetic energy tensor. 
The principal values of this tensor 
change in opposite directions when ice melts, leading to a very small net quantum mechanical 
effect on the melting point. This paper presents the first direct experimental observation of this 
phenomenon, achieved by measuring the deuterium momentum distributions $n({\bf p})$ in heavy water and ice
using Deep Inelastic Neutron Scattering (DINS), and resolving their anisotropy. 
Results from the experiments, supplemented by a theoretical analysis, 
show that the anisotropy of the quantum kinetic energy tensor can also be captured
for heavier atoms such as oxygen.
\end{abstract}

\maketitle

\begin{center}
\includegraphics[width=5cm]{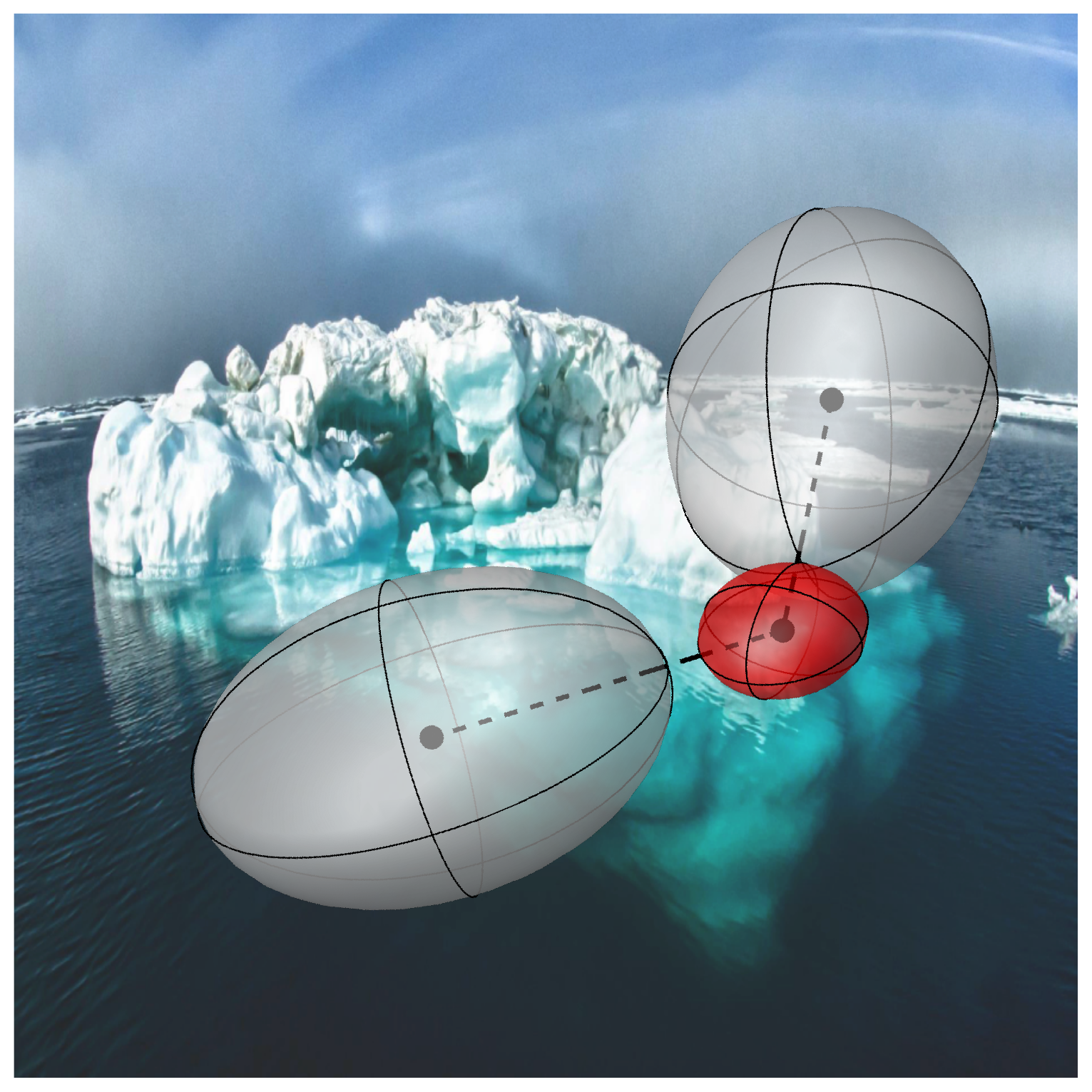} \\
{\small The iceberg image was used with permission of the NOAAr's National Ocean Servicer, 2012
(http://commons.wikimedia.org/wiki/File:Iceberg\_-\_NOAA.jpg)}
\end{center}
\noindent Adapted (pre-print) with permission from J. Phys. Chem. Lett., 2013, 4 (19), pp 3251-3256
(2013) DOI: 10.1021/jz401538r. Copyright (2013) American Chemical Society.\\
{\em keywords:} deep inelastic neutron scattering; ab initio path integral molecular dynamics; particle momentum distribution
\\

The structure and dynamics of liquid water are directly influenced by quantum mechanics, not 
only in terms of the electronic structure and chemical bonding, but also at the level of the 
nuclear motion. So-called nuclear quantum effects (NQEs) include zero-point energy, 
tunnelling, isotope effects in the thermodynamic properties, and 
-- what is most relevant to the present work -- large deviations from the classical, 
Maxwell-Boltzmann behaviour of both the 
average nuclear kinetic energy $\avg{E_K}$ and the momentum distribution $n({\bf p})$.

Even though NQEs are very large -- the zero-point energy content of an O--H stretching 
vibration is in excess of 200~meV -- it is often the case that their net effect on macroscopic 
properties is relatively small. For instance, the melting temperatures of light and heavy
water differ by less than 4~K, and the boiling temperatures by just 1~K. 
Recent theoretical analyses \cite{habe+09jcp,li+11pnas} have suggested that this 
could stem from a partial cancellation between quantum effects in the intra and inter
molecular components of the hydrogen bond -- so that the net effect is small even if 
the individual contributions are large. 
In particular, the competition between
quantum effects can be seen very clearly when decomposing the changes in quantum kinetic
energy of protons and deuterons along different molecular axes \mbox{\cite{mark-bern12pnas,liu+13jpcc}}.

The mechanism that underlies the competition between changes
in the different components of the quantum kinetic energy
can be understood by considering as an analogy a two-level 
quantum system with an environment-dependent off-diagonal coupling $\beta$. 
A small change in the coupling 
$\Delta\beta$ -- arising from a phase transition or some other change in the
environment of the system  -- 
will shift its eigenvalues by the same amount proportional to $\Delta\beta$,
{\em but in opposite directions}.
Even though this picture is clearly over-simplified, it is consistent with a diabatic state
model of the hydrogen bond,\cite{mcke12cpl} it demonstrates that the notion
of competing quantum effects is nothing exotic, and explains why it returns in many 
circumstances in the study of water and other hydrogen-bonded systems. 

Competing quantum effects have in fact been identified in a diverse variety of 
simulations \cite{habe+09jcp,li+11pnas,mark-bern12pnas,liu+13jpcc}, 
and it seems entirely plausible that they are 
at the root of the explanation for why many of the properties of water depend only
weakly on isotopic composition. 
As discussed in the Supporting Information (SI), the change of many thermodynamic 
properties on isotopic substitution can be related to changes in the quantum 
kinetic energy.~\cite{ramirez2011kinetic,mark-bern12pnas}
Under a few simplifying assumptions, one can for instance relate the change in 
the quantum kinetic energy of the D atoms when heavy water 
melts, $\Delta_{{\rm fus}}E_{\rm K}(m_{\rm D},T_{{\rm fus}}(m_{\rm D}))$, to 
macroscopic thermodynamic properties such as the entropy of melting
of light water, $\Delta_{{\rm fus}}S(m_{\rm H})$, and the change in the melting 
temperature upon isotopic substitution:
\begin{equation}
\Delta_{{\rm fus}}E_{\rm K}(m_{\rm D},T_{{\rm fus}}(m_{\rm D}))
\approx\frac{\Delta_{{\rm fus}}S(m_{\rm H}) }
{2 \left( {\sqrt{m_{\rm D}/m_{\rm H}}}-1 \right) }
 \left[T_{{\rm fus}}(m_{\rm H})-T_{{\rm fus}}(m_{\rm D}) \right]. 
 \label{eq:deltae-melt}
\end{equation}
Eq.~(1) predicts a change in kinetic energy per 
D atom of $\Delta_{{\rm fus}}E_{\rm K}=-0.5$~meV. This is a tiny value -- less than
0.5\%~ of the total kinetic energy of the D atoms at room temperature.
A direct experimental investigation of the competing
quantum effects at play here should reveal whether this
happens because of a cancellation, or because the environment of a D atom 
changes very little on melting.

Deep Inelastic Neutron Scattering (DINS), or Neutron Compton Scattering (NCS)
at high momentum and energy transfers ($\hbar q$ and $\hbar \omega$, respectively),
is an experimental technique that is particularly well suited to probe 
the quantum behavior of atomic nuclei, by directly measuring
$n({\bf p})$ \cite{andre+05advp,2011APRSphysrep,reiter}.
The results from DINS experiments have stimulated
the development of improved theoretical methods for evaluating the proton 
momentum distribution \cite{ceri+09prl2,lin+10prl,ceri+11jcp,ceri-mano12prl}, as well as their 
application to benchmark systems, with a close interplay between theory and experiment 
\cite{morr-car08prl,ceri+10prb}.
One can infer the anisotropy of the particle momentum distribution from DINS experiments 
even in cases when only the spherically averaged $n(p)$ is available.
In water, this provides insight into the local environment of the proton and can help elucidate
the nature of hydrogen bonding \cite{reit+04bjp,andre+05advp,7}, 
the structure of hydration shells, 
and the effects of confinement \cite{52,senesi,2010reiterPRL}.
Indeed this information  can be seen as the direct experimental counterpart of 
the decomposition of the quantum kinetic energy along molecular axes, which has been used 
so successfully to unravel competing quantum effects in simulations \cite{mark-bern12pnas,ceri-mano12prl,liu+13jpcc}.

The focus of DINS studies has recently broadened to consider also heavier 
atoms \cite{krzy-fern11prb,seel+12jpcm,ceri-mano12prl},
which, although challenging because of their less-pronounced quantum nature, promise a 
more comprehensive picture of the underlying physics. 
Theoretical calculations \cite{lin+11prb,2011Herrero} demonstrate a sizeable excess of kinetic 
energy for the oxygen atoms in ice, relative to the classical value.
This kinetic energy excess shows a clear dependence on the chemical 
environment \cite{seel1}, and on the microscopic structure.
A direct, accurate measurement of the kinetic energy of the oxygen atoms 
could for instance shed light on recent findings that indicate an increased 
localisation of the oxygen in heavy water compared to light water,  
as evidenced by a 10 per cent overstructuring in the heavy water $g_\text{OO}(r)$ 
radial distribution function \cite{soper}.

At present, the only instrumentation suitable to perform
measurements of $n({\bf p})$ in condensed matter systems
is the VESUVIO spectrometer, which operates on a dedicated beam line
at the pulsed neutron source ISIS (Rutherford Appleton Laboratory, UK) \cite{Vesuvio,reiter}.
The instrument uses  neutrons with incident energies in the 
range 1-800 eV,
and relies on the fact that, at sufficiently high momentum transfers, 
any scattering process can be described within the Impulse Approximation (IA) \cite{west}. 
This implies that the neutron scatters from a single atom, with conservation 
of the total kinetic energy and momentum of the neutron and the atom \cite{gunn}. 
In the IA regime the inelastic neutron scattering cross section is related in 
a simple way to $n({\bf p})$. The neutron scattering function $S_{\rm IA}(\bf q,\omega)$ is
\begin{equation} \label{sia}
\frac{\hbar {\it q}}{\it m} S_{\rm IA}(\bf q,\omega)= \it J_{\rm IA}(y, \hat{\bf q})=
\int n({\bf p}) \delta\left(y-{\bf p \cdot \hat q}\right) \mathrm{d} {\bf p}
\end{equation}
where $(\bf q,\omega)$ are the wave vector and energy transfers,  
$m$ is the mass of the atom being struck,  
$y=\frac{m}{\hbar q} \left[\omega -\frac{\hbar q^2}{2m}\right]$ is 
the particle momentum  along the $\hat{\bf q}$ direction, 
and $J_{\rm IA}(y,\hat{\mathbf{q}})$ is the neutron Compton Profile (NCP)
\footnote[0]{For consistency
with previous literature and ease of notation we write the momentum as a wave vector.}.

When the sample is isotropic, the particle momentum distribution only depends on the
modulus of ${\bf p}$, and the $\hat{\bf q}$ direction is immaterial, so the NCP
is simply $J_{\rm IA}(y)=2\pi\int_{\left|y\right|}^\infty pn(p)\mathrm{d}p$.
This ideal peak profile is broadened by finite-$q$ correction terms $\Delta J(y,q)$ 
\rome{as discussed in the SI,~\cite{sears}}
and by convolution with the 
instrumental resolution function $R(y,q)$, so the experimental NCP, $F(y,q)$, is
\begin{equation}
F(y,q)= [J_{\rm IA} (y) + \Delta J(y, q)] \star R(y,q).
\end{equation}

One reasonable (and also insightful) way to extract the
physical information content from the experimental $F(y,q)$ profile is to 
assume that the underlying $n(p)$ arises from the spherical average of an anisotropic Gaussian distribution 
\cite{ceri+10prb,lin+11prb,and,2011Flammini},
\begin{equation} \label{ndp} 
4\pi p^2  n(p)=\int  \frac{\delta(p-|\mathbf{p}|)}{\sqrt{8 \pi^3}\sigma_x \sigma_y \sigma_z} \exp\left(-
\frac{p_x^2}{2\sigma_x^2}-\frac{p_y^2}{2\sigma_y^2}-\frac{p_z^2}{2\sigma_z^2}\right)\,{\rm d}^3{\bf p}.
\end{equation} 
This expression involves three  parameters -- the variances $\sigma^2_{\alpha}$ for $\alpha=x,y,z$ -- which 
are related to three effective principal frequencies $\omega_{\alpha}$ by 
$ \sigma_{\alpha}^2 =\frac{m\omega_\alpha}{2\hbar}\coth\frac{\beta\hbar\omega_\alpha}{2}$, 
or to the three components 
of the quantum kinetic energy by $\avg{E_\alpha}=\hbar^2\sigma_{\alpha}^2/2m$.
In the present study, this approach has been used to interpret  DINS data acquired on 
heavy water in the solid at 274 K, and in the liquid at 280~K and 300~K. 

To complement this experimental study, we have also performed some new 
{\em ab initio} computer simulations of heavy water and ice, using the same 
density functional theory (DFT) framework \cite{lee+88prb,beck88pra,goed+96prb,vand-krac05cpc} 
as described in Ref.~\citenum{ceri-mano12prl}. Tests with different basis sets and the inclusion
of dispersion corrections produced no qualitative changes in the results.
Nuclear quantum effects were incorporated using the PIGLET 
technique \cite{ceri-mano12prl}, which combines the path integral 
formalism \cite{feyn-hibb65book,cepe95rmp} with a correlated-noise Langevin equation
\cite{ceri+11jcp,ceri-mano12prl},
thereby enabling fully converged results for room-temperature water to be obtained 
with as few as six path integral beads. 

The conventional way to extract the particle momentum distribution from the
path integral formalism involves opening the path and is computationally 
very demanding \cite{morr-car08prl}.
A simpler alternative is to assume that the momentum distribution can be modelled as
a multivariate Gaussian as in (4), and to use the eigenvalues of the quantum kinetic 
energy tensor $\hbar^2\avg{p_{\alpha}p_{\beta}}/2m$  to estimate the principal components of this
distribution. The only difficulty here lies in the fact that in the liquid the orientations 
of the water molecules change with time, so one cannot simply average the centroid 
virial estimator to obtain the anisotropic kinetic energy tensor. 

Here we compare two different ways around this difficulty. One is to perform
a running average of the kinetic energy estimator\cite{ceri-mano12prl} 
-- the so-called ``transient anisotropic Gaussian'' (TAG) approximation.
For this we used a triangular averaging window of 100~fs, which has previously been 
shown to give converged
results for light water.\cite{ceri-mano12prl}. Another possibility is to assume that 
the principal axes of the kinetic energy tensor will have a fixed orientation relative 
to the molecular geometry. One can then perform a mean-square displacement (MSD) 
alignment of the instantaneous configuration of each water molecule to a reference 
structure, rotating the kinetic energy estimator into the molecular reference frame, 
and computing its average and its eigenvalues \cite{mark-bern12pnas,liu+13jpcc}.
We will show that the two approaches give results that are consistent with one another,
and that they enable a direct comparison with the DINS experiment.

\begin{figure}
\centerline{\includegraphics*[width=0.95\columnwidth]{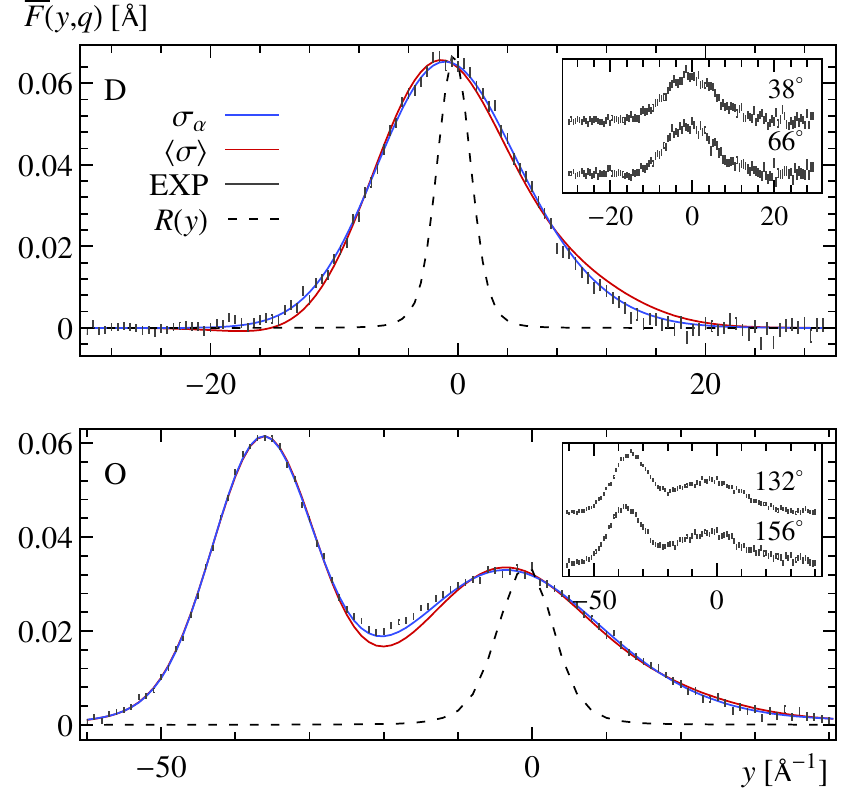}}

\caption{\label{profiles}
(Color online) Experimental  NCS profiles for heavy ice at T= 274 K. The two panels
reports the detector-averaged NCS for D (upper panel) and O (lower panel). 
Best fits using simple Gaussian  (red line) and spherically averaged 
multivariate Gaussian (blue line) approximations to the momentum distribution
are also reported. In the case of the O profiles, 
the  peak at $y=-40$ \AA$^{-1}$ is due to the contribution from the Cu 
sample container.  The instrumental resolution is reported as a black dotted line, and 
the insets show examples of the raw data from two individual detectors for D 
and from two groups of detectors at the same scattering angles for O.  }
\end{figure}

Figure~\ref{profiles} reports examples of the experimental
detector-averaged NCP for D and O, $\bar F(y, q)$, together with the 
best fits obtained with an isotropic and a multivariate Gaussian ansatz for $n(p)$.  \rome{
The angle-averaged $\bar F(y, q)$ is obtained by averaging over the detectors 
in the range 32$^{\circ}$ to 66$^{\circ}$ for D and those between 130$^{\circ}$ and 163$^{\circ}$ for O. 
Data were not symmetrized. This figure provides a graphical representation of the 
overall quality of both the data and the fit (see also the SI). 
}
Clearly, the multivariate Gaussian profile provides a better fit to the 
experimental data than an isotropic Gaussian. 

\begin{figure}
\centerline{\includegraphics*[width=0.95\columnwidth]{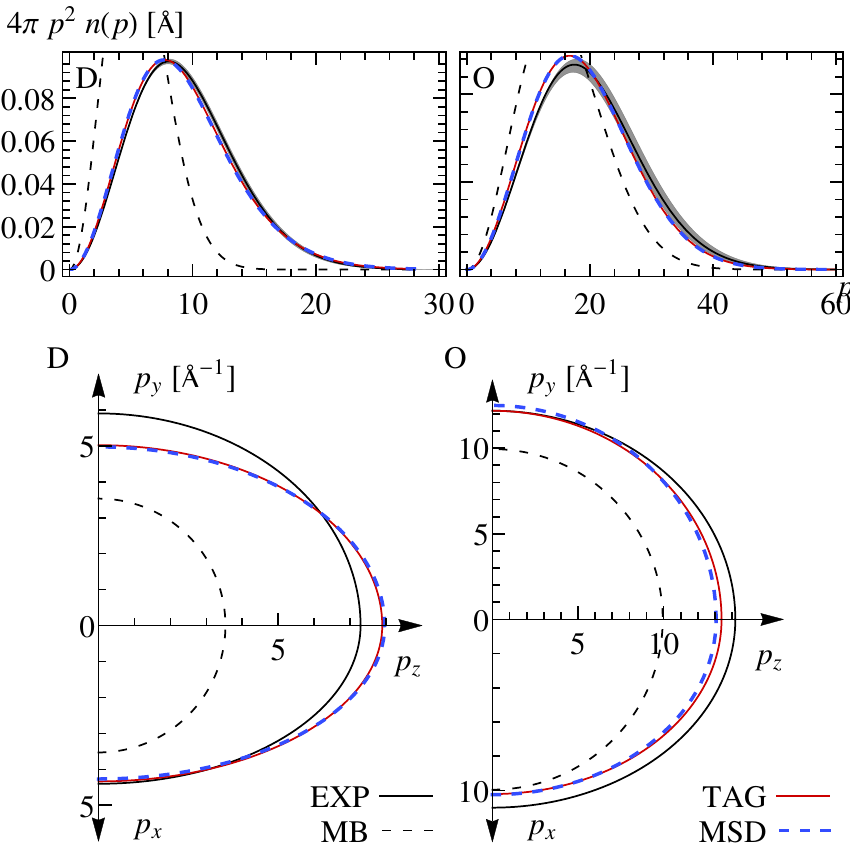}}
\caption{\label{pplots} (Color online) Comparison of the
momentum distributions of O and D in liquid D$_2$O at 300~K, as obtained from the 
analysis of DINS experimental data and from two different analyses of the 
\emph{ab initio} PIGLET simulations.  In all plots, the continuous black
curve corresponds to the experimental data, the red curve to a TAG analysis and the blue
curve to a MSD analysis of the simulation data. The dashed line corresponds to 
what would be expected if $n(p)$ were just a classical Maxwell-Boltzmann
distribution. The upper panels show the spherically averaged $n(p)$ 
\rome{(the shaded area around the experimental curve represents the 
confidence interval)}, while the 
lower panels contain a graphical representation of the anisotropy. The curves 
correspond to iso-surfaces of the $n({\bf p})$ cut along the $xz$ and $yz$ planes. 
The contour line is chosen in such a way that the intercepts on the axes are the values of $\sigma_\alpha$. }
\end{figure}

In the upper panels of Figure~\ref{pplots}, we compare the experimental and theoretical $n(p)$s for 
liquid D$_2$O at 300~K. There is a near-perfect agreement between theory and experiment in 
the case of D. The discrepancy is larger in the case of O, but \rome{comparable with the error
bar and} much smaller than the deviation from a classical, Maxwell-Boltzmann distribution. 

The lower panels show that the discrepancy between theory and experiment 
is more pronounced when one focuses on the anisotropy of the distribution. 
The TAG and MSD approximations are consistent with each other, as discussed
in more detail in the SI. It is interesting that, 
despite the noticeable differences in the individual values of $\avg{E_{\alpha}}$, the theoretical and 
experimental $n(p)$s for D are almost indistinguishable. 
However, the theoretical and experimental $n(p)$s for O, 
which involve a larger discrepancy in the total $\avg{E_K}$ but smaller 
discrepancies in the individual components, show a more evident difference.
Because of the averaging in Eq.~(4),
the computed $n(p)$ depends only weakly on how the kinetic energy components
are distributed, but in a more pronounced way on the total kinetic energy.

The relative insensitivity of $n(p)$ to the partitioning of $\avg{E_K}$ into three
principal components justifies the use of either the TAG or the MSD approach to 
estimate the anisotropy of the kinetic energy tensor. 
However, this insensitivity also means that extracting the anisotropy from the 
spherically-averaged $n(p)$ is an ill-conditioned problem. 
For this reason, the analysis of the experimental data
typically yields larger relative errors in the individual components of the 
kinetic energy than in the total.

\begin{table}[htbph!]
\caption{\label{comparison} Comparison between theoretical and experimental 
components of the quantum kinetic energy for D and O in heavy water, at different 
temperatures. All values are in meV, and the theoretical results have a statistical error bar
smaller than 0.1~meV. We also report the computed center-of-mass mean kinetic energy
$\avg{E_{COM}}$ of the D$_2$O molecules.}
\centering
\begin{tabular}{ c c c c c }
\hline\hline
                     &  D [exp] & D [TAG/MSD]   &O [exp]  &O [TAG/MSD]  \\ \hline
\multicolumn{5}{l}{D$_2$O, $T=300$~K, liquid \hfill $\left<E_{COM}\right>=42.1$ }           \\ \hline
$\left<E_x\right>$   & 20.1$\pm$1.1 &  19.5  / 18.9    &  15.8$\pm$1.7  &  13.6 / 13.7      \\       
$\left<E_y\right>$   & 36.1$\pm$2.3 &  26.1  / 25.6    &  19.5$\pm$1.3  &  19.4 / 20.4      \\       
$\left<E_z\right>$   & 55.1$\pm$2.3 &  64.6  / 65.7    &  26.3$\pm$1.5  &  23.4 / 22.3      \\       
$\left<E_K\right>$   &111.3$\pm$3   & 110.2            &  61.6$\pm$3.1  &  56.4             \\       
\hline
\multicolumn{5}{l}{D$_2$O, $T=280$~K, liquid \hfill $\left<E_{COM}\right>=39.5$}            \\ \hline
$\left<E_x\right>$   & 18.8$\pm$1.1 &  19.4  / 18.9    &  16.0$\pm$2.3  &  13.6 /  13.7     \\       
$\left<E_y\right>$   & 38.6$\pm$2.5 &  25.7  / 25.2    &  21.0$\pm$0.6  &  19.2 /  20.2     \\       
$\left<E_z\right>$   & 54.2$\pm$2.4 &  63.6  / 64.6    &  24.1$\pm$2.1  &  23.2 /  22.2     \\       
$\left<E_K\right>$   & 111.6$\pm$2  &  108.7           &  61.1$\pm$3.1  &  56.1             \\       
\hline
\multicolumn{5}{l}{D$_2$O, $T=274$~K, liquid \hfill $\left<E_{COM}\right>=38.9$}            \\ \hline
$\left<E_x\right>$   &              &  19.3  / 19.0    &                &  13.4 / 13.5      \\       
$\left<E_y\right>$   &              &  25.8  / 25.3    &                &  19.1 / 20.1      \\       
$\left<E_z\right>$   &              &  63.2  / 64.1    &                &  23.1 / 22.0      \\       
$\left<E_K\right>$   &              &  108.3           &                &  55.6             \\       
\hline
\multicolumn{5}{l}{D$_2$O, $T=274$~K, solid \hfill $\left<E_{COM}\right>=39.2$ }            \\ \hline
$\left<E_x\right>$   & 22.5$\pm$1.8 &  20.1  / 19.8    &   16.1$\pm$2.3 &  13.7 / 13.8      \\       
$\left<E_y\right>$   & 37.4$\pm$2.5 &  26.3  / 25.9    &   20.1$\pm$1.6 &  19.0 / 19.9      \\       
$\left<E_z\right>$   & 48.1$\pm$3.4 &  61.9  / 62.4    &   24.2$\pm$1.4 &  23.0 / 21.9      \\       
$\left<E_K\right>$   & 108.0$\pm$2  &  108.3           &   60.4$\pm$4   &  55.7             \\    
\hline\hline
\end{tabular}
\end{table}

Bearing this in mind, let us now discuss how DINS can 
provide a direct verification of the concept of competing 
quantum effects in water. Table~\ref{comparison} 
collects all of the present experimental and theoretical results together in a compact form. 
The agreement between the total deuterium kinetic energy obtained by DINS and 
by simulation is almost perfect. The change in kinetic energy between the liquid
at 300~K and 280~K is much smaller than the drop in classical thermal energy, 
which is consistent with the deuteron being almost completely frozen in 
its vibrational ground state. There is also agreement with previous experiments 
at $T=292$~K within their (much larger) error bar~\cite{giuliani2011isotope}.

The most interesting results in Table~\ref{comparison} concern the 
behaviour of the momentum distribution in heavy water upon freezing.
When going from the liquid to the solid, the DINS data show substantial changes in
$\avg{E_x}_\text{D}$ (associated with motion perpendicular
to the plane of the water molecule~\cite{mark-bern12pnas,ceri-mano12prl}) 
and in $\avg{E_z}_\text{D}$ (associated with motion parallel to the covalent O--H bond). 
However, the two components change in opposite directions, leading to a much smaller
change in the total kinetic energy, which is not statistically significant given the 
experimental error bars. 
The increase in $\avg{E_x}_\text{D}$ is a signature of the more hindered librations
in the solid phase, while the decrease
of $\avg{E_z}_\text{D}$ signals a weakening of the covalent bond, which is consistent with 
the red shift of the stretching peak observed in the vibrational spectroscopy of ice~\cite{eise-kauz68book}.
These observations therefore provide a direct experimental verification of the 
competition between quantum effects resolved along different molecular axes.

Simulations predict the same qualitative effect on the different components
of $\avg{E_K}$:  $\avg{E_z}_\text{D}$ decreases on freezing, 
but $\avg{E_x}_\text{D}$ increases, leaving almost no change in the 
total kinetic energy. Performing simulations of the liquid at 280K, and 
of both the liquid and the solid at 274~K, allows us to infer that these 
effects are due to the phase transition and not the 6~K temperature drop. 
Note that our simulations show no sign of an increase in quantum kinetic energy upon supercooling,
confirming previous theoretical results for light water~\cite{ramirez2011kinetic}.
The present experiments were deliberately performed well into the stable solid and liquid 
phases of heavy water, in order to focus on the experimental signature of competing
quantum effects on melting without interference from the more controversial effects that
have been observed in DINS measurements on supercooled water~\cite{6,giuliani2011isotope}. 

While experiment and theory agree on the qualitative observation of a competition
between quantum effects on melting, there are quantitative discrepancies that deserve
further comment. For one thing, our DFT results predict $\Delta_\mathrm{fus}E_K\approx 0$,
whereas simple thermodynamic arguments predict $\Delta_\mathrm{fus}E_K\approx -0.5$meV 
(see Eq.~(1)). 
As discussed in the SI, a simple, empirical water model~\cite{habe+09jcp} yields 
predictions that are in agreement with the macroscopic thermodynamic data -- which
is perhaps unsurprising given that this empirical model accurately describes 
the change in melting temperature upon isotope substitution\cite{rami-herr10jcp}. 
While it is remarkable that an {\em ab initio} 
calculation can get so close to the correct result, it is clear that
DFT has not yet reached the level of accuracy necessary to obtain a quantitative description of
isotope effects. Neither have the DINS experiments reached the exquisite
level of accuracy that is necessary to discern such a minute change in the total kinetic energy.
\rome{Currently the overall sensitivity of DINS measurements allows one to 
infer values for proton $\avg{E_k}$ within 2 meV uncertainty,~\cite{flammini2012spherical}
and similar if not higher uncertainty for the heavier masses, D and O. 
Although changes of the order of 0.5 meV are beyond the current sensitivity of the instrument, 
we have demonstrated} that one can nevertheless gain insight into the competition of
effects that leads to a small kinetic energy change by resolving the anisotropy 
of the kinetic energy tensor. The quantitative differences between DINS and PIGLET on
the individual components of the kinetic energy, however, indicate that at present this insight
is only qualitative.

Table~\ref{comparison} also presents the results for the oxygen momentum distribution. 
While there is good qualitative agreement between theory and experiment, we
observe a discrepancy of almost 10\% in the total kinetic energy, which may 
stem from shortcomings of the 
modelling or from the analysis of the experimental data --
which is made harder by the weaker signal given by oxygen and 
by the partial overlap between the $F(y,q)$ peak of the O and that of the Cu can. 
Nevertheless, the analysis  captures even the comparatively weak 
anisotropy of the oxygen atom kinetic energy, demonstrating how promising 
it is to extend DINS to heavy atoms. We anticipate that analysis,~\cite{blostein2005formalism} 
software, and 
instrument upgrades planned on VESUVIO in the near future will enable
a greater precision in the simultaneous measurement of light and heavy atoms,
enabling one to access quantitative as well as qualitative information 
on the particle momentum distribution.

This work was partially supported within the CNR-STFC Agreement 
No. 06/20018 concerning collaboration in scientific research at 
the spallation neutron source ISIS. CA and RS acknowledge the support of META 
(Materials Enhancement for Technological Applications) Marie Curie Actions, 
People, FP7, PIRSES-GA-2010-269182.
MC acknowledges funding from 
the EU Marie Curie IEF No. PIEFGA-2010-272402 and computer time from CSCS 
(project ID s388).
DEM acknowledges funding from the Wolfson Foundation and the Royal Society.

\paragraph*{Supporting Information Available:}
Additional details on the relation between the change in kinetic energy upon melting
and macroscopic thermodynamic observables, on the analysis of the experimental data, on
the technical aspects of the simulations, and on the results obtained with an empirical
force field. This material is available free of charge via the Internet \emph{http://pubs.acs.org}.

\providecommand*{\mcitethebibliography}{\thebibliography}
\csname @ifundefined\endcsname{endmcitethebibliography}
{\let\endmcitethebibliography\endthebibliography}{}

\newpage

\end{document}